# Observation of topological transition of Fermi surface from a spindle-torus to a torus in large bulk Rashba spin-split BiTeCl

Fei-Xiang Xiang[1], Xiao-Lin Wang[1]*, Menno Veldhorst[2], Shi-Xue Dou[1], Michael S. Fuhrer[3]

[1]*Institute for Superconducting and Electronic Materials, Australian Institute for Innovative Materials, University of Wollongong, Innovation Campus, North Wollongong, New South Wales 2500, Australia,*

[2]*Centre for Quantum Computation and Communication Technology, School of Electrical Engineering and Telecommunications, The University of New South Wales, Sydney, New South Wales 2052, Australia,*

[3]*School of Physics, Monash University, Clayton, Victoria 3800, Australia*

*Corresponding author: xiaolin@uow.edu.au*

**The recently observed large Rashba-type spin splitting in the BiTeX (X = I, Br, Cl) bulk states due to the absence of inversion asymmetry and large charge polarity enables observation of the transition in Fermi surface topology from spindle-torus to torus with varying the carrier density. These BiTeX systems with high spin-orbit energy scales offer an ideal platform for achieving practical spintronic applications and realizing non-trivial phenomena such as topological superconductivity and Majorana fermions. Here we use Shubnikov-de Haas oscillations to investigate the electronic structure of the bulk conduction band of BiTeCl single crystals with different carrier densities. We observe the topological transition of the Fermi surface (FS) from a spindle-torus to a torus. The Landau level fan diagram reveals the expected non-trivial $\pi$ Berry phase for both the inner and outer FSs. Angle-dependent oscillation measurements reveal three-dimensional FS topology when the Fermi level lies in the vicinity of the Dirac point. All the observations are consistent with large Rashba spin-orbit splitting in the bulk conduction band.**

Spin-orbit coupling (SOC) is a relativistic effect present in a system with broken symmetry, where a charged particle moving in an electric field experiences an effective magnetic field, which interacts with its spin[1,2]. In solids, electrons move in a crystal potential, and if there is a potential gradient, effective SOC arises[2] and manifests itself in the spin-split band structure. Such spin-splitting was firstly described by Dresselhaus[3] and Rashba[4] model in zinc blend and wurtzite structure, respectively, and later by Bychkov-Rashba model[5] at surface and interface. Although large spin splitting has been observed at the surface of heavy metals, it remains small in conventional semiconductors.

Recently large Rashba-type spin splitting has been observed in the bulk bands of polar semiconductors BiTeX (X = I, Br, Cl) due to the broken inversion symmetry and charge polarity in the bulk[6-11]. The high energy scale of the Rashba effect in BiTeX provides opportunities for achieving practical spintronic applications[12,13] and realizing non-trivial phenomena such as the intrinsic spin Hall effect[14], non-centrosymmetric exotic superconductivity[15,16], Majorana fermions[17,18], and topological transitions of the Fermi surface (FS)[9,19,20]. Among them, observation of the topological transition of FS from a spindle-torus to a torus is an important step toward exploring spin-dependent transport and other exotic physical phenomena in the low-carrier-density regime[9,19,20]. In particular, it is highly desirable to tune the Fermi level into the vicinity of the band-crossing point (Dirac point) at zero momentum in the Rashba system, since schemes to realize Majorana fermions[18] involve opening a small energy gap at the Dirac point to realize a single spin non-degenerate band. Furthermore, the spin polarization of current is largest in the vicinity of the Dirac point[20,21], hence such materials could be used as spin injectors as has been proposed with topological insulators[22,23], but in this case the effect would not require surface-dominated transport.

Magnetotransport is a powerful method to study the electronic properties in materials such as topological insultors[24,25]. In particular, Shubnikov-de Haas (SdH) oscillation can probe the electronic structure, reveal information on the Fermi surface (FS) topology[26,27], and access the Berry phase[25,28-30], so it is highly suitable for investigating topological transitions of the Fermi surface and detecting the potential topological surface states. Although BiTeCl has smaller spin splitting than BiTeI, it exhibits a larger band gap and more isotropic spin splitting[10] which are very desirable for transport measurement. In this paper, using Shubnikov-de Haas effect we observe topological transition of FS from spindle-torus to torus in BiTeCl single crystals at various carrier densities. The Landau level fan diagram reveals the nontrivial $\pi$ Berry phase both in the inner FS (IFS) and the outer FS (OFS). We also resolve three-dimensional FS topology when Fermi level lies in the vicinity of Dirac point by angle-dependent oscillation measurement. All the observations are consistent with large Rashba spin-orbit splitting in the conduction band of BiTeCl.

## Results

**Observation of the topological transition of Fermi surface**. Contrasting with BiTeI, the Rashba spin-orbit splitting in BiTeCl occurs in the $\Gamma K$-$\Gamma M$ plane of momentum space, denoted as the $k_\parallel$ plane, where momentum along the $z$ direction, $k_z = 0$[10,11,31,32]. The energy-momentum dispersion can be described by the following equation, assuming a parabolic band.

$$E_\pm(k) = \frac{\hbar^2}{2m^*}(k \pm k_0)^2 \qquad (1)$$

where $k = \sqrt{k_x^2 + k_y^2}$, $k_0$ is the momentum offset caused by the Rashba spin splitting, and $m^*$ is the effective mass of the electrons and $\hbar$ Planck's constant divided by $2\pi$. Besides the $k_0$, the other two Rashba parameters are the Rashba energy, $E_R = \hbar^2 k_0^2/2m^*$, and the Rashba constant, $\alpha_R = 2E_R/k_0$, which represent the energy when $k = 0$ and the strength of the Rashba effect, respectively. The spin-split conduction band dispersion of BiTeCl near the $\Gamma$ point is shown in Fig. 1a, the insets of which illustrate the cross section of FS at $k_\parallel$ plane when the Fermi energy $E_F$ has different values. When $E_F > E_R$, the three-dimensional FS is a spindle-torus and both the IFS (spindle) and OFS (torus) are present [inset (i) of Fig. 1a]. While in a small Rashba spin-split system, the IFS and OFS result in beating patterns in the SdH oscillations; in a giant Rashba spin-split system, the two sets of oscillations are thoroughly decoupled from each other[29]. When $E_F \lesssim E_R$, the IFS vanishes and the FS is simply a torus, i.e. only the OFS is present [inset (ii) and (iii) of Fig. 1a]. Thus it is expected that the two sets of oscillations represent a transition to a single frequency in the Shubnikov-de Haas oscillation measurement when the Fermi level approaches the Dirac point, corresponding to the topological transition of FS from spindle-torus to torus.

BiTeCl is a degenerate semiconductor due to the self-doping effect (nonstoichiometric effect or formation of defects) which is similar to the topological insulators such as $Bi_2Se_3$ and $Bi_2Te_3$. Because the self-doping effect depends on the temperature gradient along the quartz tube during the single crystal growth, the carrier density of crystal can vary in different positions in the quartz tube. To observe the expected topological transition of the FS as the Fermi level approaches the Dirac point, a group of single crystal samples with various carrier concentrations was selected from different position in the tube. The samples are denoted as S1 to S7 ordered according to increasing OFS oscillation frequency. Figure 1b,c shows the typical longitudinal resistivity ($\rho_{xx}$) and Hall resistivity ($\rho_{xy}$) of the samples. The negative and linear slope indicates that the dominant carriers are electrons. The calculated carrier density is $8.87 \times 10^{18}$ cm$^{-3}$. A clear evolution of the transition from two sets of oscillation to a single frequency oscillation is shown in Fig. 1d that plots the first derivative of the longitudinal resistance, $R_{xx}$, versus $B$ for samples S1 to S7. From S7 to S1 the oscillations from IFS indicated by the red arrows gradually disappear and only oscillations from OFS are left, consistent with the topological transition of the FS described above[9,19].

**Standard Shubnikov-de Haas oscillation analysis**. To deduce the electronic structure via standard SdH oscillation analysis, the Lifshitz-Kosovich (LK) formula is used as follows[26,27,29]:

$$\frac{\Delta \rho}{\rho_0} = \frac{5}{2}\left(\frac{B}{2F}\right)^{\frac{1}{2}} \frac{2\pi^2 k_B T m^*/\hbar eB}{\sinh(2\pi^2 k_B T m^*/\hbar eB)} e^{-\frac{2\pi^2 k_B T_D m^*}{\hbar eB}} \cos\left(\frac{F}{B} + \frac{1}{2} - \frac{\Phi_B}{2\pi} + \delta\right) \qquad (2)$$

where $F$ is the oscillation frequency, $k_B$ the Boltzman constant, $e$ the elementary charge, $T$ the temperature, $T_D$ the Dingle temperature, $\Phi_B$ the Berry phase, and $\delta$ the phase shift determined by the dimensionality. Figure 2a shows OFS SdH oscillations of S2 at various temperatures after subtracting the background. Fitting the temperature dependence of the OFS oscillation amplitudes of S2 around 11.65 T with the thermal damping factor $\frac{2\pi^2 k_B T m^*/\hbar eB}{\sinh(2\pi^2 k_B T m^*/\hbar eB)}$ from Equation (1), as shown in Fig. 2b, yields the effective mass, $m^* = 0.191 \pm 0.005\ m_e$, where $m_e$ is the free electron mass. Because $\rho_{xx} < \rho_{xy}$, as shown in Fig. 1b,c, the minima and maxima of the SdH oscillations are assigned as integer ($n$) and half integer ($n+1/2$) LL indices, respectively. Linear fits of the Landau level (LL) indices vs. 1/$B$ of S1 to S7 in the LL fan diagram of Fig. 3c yield the slopes of the fit lines and their intercepts with the LL index axis which correspond to oscillation frequency $F$ and phase factor $-\frac{\Phi_B}{2\pi} + \delta$, respectively. Taking the value $\delta = \pm\frac{1}{8}$ for our three-dimensional (3D) system[26,27,29], the range of the intercepts from 0.375 to 0.625 indicates a nontrivial $\pi$ Berry phase. The inset of Fig. 2c shows the Berry phase for S1 to S7 and nearly all of them are located in the range from 0.375 to 0.625. After obtaining the effective mass, the oscillation frequency, and the phase factor, we fit the OFS SdH oscillation data of S2 with Equation (2) to obtain the Dingle temperature $T_D$. As shown in Fig. 2d, the LK formula fits well with the experimental data, resulting in $T_D = 17.6 \pm 0.12$ K. The same analysis process was applied to S1 and S3 to S6 (see Part A of Supplementary Information). Only the oscillation frequency was obtained for S7, however. All of the fitting parameters, $F$, $m^*$, and $T_D$, are tabulated in Table 1.

Figure 3a,b shows $-d^2R_{xx}/dB^2$ as a function of magnetic field for S5 and S7, the maxima and minima of which correspond to the oscillation maxima and minima. Besides the high frequency oscillations from the OFS in high field, low frequency oscillations from the IFS can be observed in low field, as indicated by the blue arrows. Figure 3c shows the LL fan diagram for the IFS. Because the SdH oscillations from IFS for S5 and S7 are very close to the regime of $\rho_{xx} \sim \rho_{xy}$, we carefully assigned the integer LL $n$ as discussed in the Part D of supplementary information. For S5 the integer LL $n$ is assigned to the maxima of SdH oscillations and for S7 the low field part the integer LL $n$ is assigned to the maxima of SdH oscillations and in the high field part the integer LL $n$ is assigned to the minima of SdH oscillations. The linear fittings yield the frequencies 2.81 ± 0.10 and 5.93 ± 0.01 T, respectively, and the intercepts with the LL index axis are at 0.756 ± 0.048 and 0.700

± 0.003, respectively, which are very close to the region from 0.375 to 0.625 and indicates that the IFS also has a non-trivial π Berry phase. The nontrivial π Berry phase in the OFS and IFS is consistent with the pure bulk Rashba effect which was first reported in BiTeI[18].

**Determination of Rashba parameters and Femi levels**. We now determine the Rashba parameters and then calculate the Fermi levels of the seven samples from the Rashba model. According to the Onsager-Lifshitz equation, $F = (\hbar/2\pi e)A_F$, where $A_F$ is the extremal area of cross section of FS perpendicular to the magnetic field direction. With $A_F = \pi k_F^2$, the Fermi wave vector is obtained, $k_F = \sqrt{2eF/\hbar}$, as given in Table 1. Substituting the IFS wave vector $k_F^{IFS}$ and the OFS wave vector $k_F^{OFS}$ of S5 and S7 into Equation (1) yields $k_0$ = 0.03158 and 0.03006 Å$^{-1}$, respectively; the similar values indicate self-consistency of the model. Then the Rashba energy, $E_R = \hbar^2 k_0^2/2m^*$ = 18.45 meV, and Rashba constant, $\alpha_R = 2E_R/k_0$ = 1.20 eVÅ, is obtained with the average of the two $k_0$ and average effective mass of S1-S6, which agree with theoretical values[31,32]. Substituting the $k_F$ of S1-S7 into Equation (1) yields the corresponding Fermi energies given in Table 1. With $k_0$ and $m^*$ determined, the dispersion relation is plotted in Fig. 3(d), and the solid lines are the calculated Fermi levels for the seven samples.

**Resolution of a 3D Fermi surface**. When the SdH oscillation with a single frequency exhibits a nontrivial π Berry phase, it is easy to relate the oscillation to the topological surface state in BiTeCl, which was observed by ARPES recently[33]. The emergence of two sets of oscillations, however, rules out this possibility in this measurement. Furthermore, as tabulated in Table 1 the carrier densities calculated from the SdH oscillations with 3D model $n = (1/3\pi^2)(2eF/\hbar)^{3/2}$ are consistent with the Hall effect measurements in the typical sample S4, $8.87 \times 10^{18}$ cm$^{-3}$, which also indicates the bulk origin of the oscillation.

To resolve the 3D FS topology, the SdH oscillations were measured over an extended range of angles with the measurement configuration shown in Fig. 4a. The oscillations exhibit symmetry in ± $\theta$ (Fig. 4b), which is corresponding to the symmetry of the FS. With increasing tilt angle |$\theta$|, the

number of the observed oscillation period becomes less and the amplitude diminishes, which is similar to the behaviour of the two-dimensional (2D) electronic system. Figure 4c plots the oscillations as a function of $1/B\cos\theta$ from 0° to 52°. From 0° to 12°, the oscillation can be reasonably described by a 2D FS, the period of oscillation depends on $1/B\cos\theta$, but as the tilt angle increases further, the oscillations deviate from the expectation for a 2D FS. While the oscillation signal cannot be extracted for angles between 56° and 82°, clear oscillations are again visible for angles around 90° with a spherical FS character, depending on $1/B$ rather than $1/B\cos\theta$, however with amplitude much smaller than those around 0° (Fig. 4d).

Now we propose a quantitative 3D FS topology which captures the above oscillation feature as shown in Fig. 4e. It is the vertical cross-section view of a spindle-torus shape FS where the black and purple solid lines denote the OFS and IFS, respectively. While the OFS deviates from the cylindrical FS shape of a strictly 2D electronic system (represented as a blue rectangle), it can be well described by a prolate spheroid (denoted as a solid red line when $\theta < 52°$). To describe the FS quantitatively, angle-dependent oscillation frequency is plotted in Fig. 4f. The experimental data (black solid circles) deviates from the 2D behaviour represented by the blue solid line but can be fit well, when $\theta < 52°$, by a prolate spheroid ($F = \frac{\hbar b^2}{2e}\sqrt{(\frac{a}{b})^2 \sin^2\theta + \cos^2\theta}$, see Part D of Supplementary Information) with a major axis a = $119.6 \times 10^{-3}$ Å$^{-1}$ and a minor axis b = $65.69 \times 10^{-3}$ Å$^{-1}$ locating on $k_z$ aix and $k_\parallel$ plane of the momentum space, respectively. The spherical FS behaviour around 90° may be caused by the hollow shape of the OFS near the zero momentum which makes the extremal cross section area constant around 90°. The half height of the OFS, c ≈ $34.5 \times 10^{-3}$ Å$^{-1}$, is estimated by treating the extremal cross section of OFS as a rectangular.

## Discussion

In this work we studied the SdH oscillation from the bulk Rashba spin-split conduction band of BiTeCl. The transition from two-frequency to single-frequency oscillation reveals the topological transition from a spindle-torus to a ring-torus FS. The momentum offset, Rashba energy, and Rashba coupling constant have been determined for giant Rashba spin splitting in BiTeCl. Both the inner and the outer Fermi surfaces have a non-trivial π Berry phase. Angle-dependent oscillation measurements reveal the three-dimensional FS topology when the Fermi level lies in the vicinity of the Dirac point.

**Methods**

**Single crystal growth.** Single crystals of BiTeCl were grown by the self-flux method according to the $Bi_2Te_3$-$BiCl_3$ binary phase diagram. $Bi_2Te_3$ was synthesized from high-purity (5N) Bi and Te powders. $Bi_2Te_3$ and $BiCl_3$ (5N) powders were weighed out with the molar ratio of 1: 9 and thoroughly ground together, which was carried out in an oxygen and moisture monitored glove box to prevent the deliquescence of $BiCl_3$. The mixture of powders then was loaded into a quartz tube and sealed under vacuum before heating to above 420–500°C over several hours. The temperature was maintained for 12 h, and then the samples were slowly cooled down to 200 °C over several days. The plate-like crystals were obtained by chemically removing the residual flux of $BiCl_3$.

**Measurement methods and conditions.** Single crystal samples with shining surfaces were cleaved from the as-grown crystals and used for standard four probe transport measurements. In some cases, a six-probe Hall measurement was employed to obtain the longitudinal resistance $R_{xx}$ and the Hall resistance $R_{xy}$ simultaneously. Gold wires were attached to the sample surface by silver epoxy, which was cured at room temperature before the measurements to ensure Ohmic contacts. Magnetic field was applied perpendicular to the sample surface up to 13.5 T, except for the angle dependent measurements. The temperature for magneto-transport measurements was 2.5 K, except for measurements of the SdH oscillations at various temperatures.

Notes: (1) Initially we observed quantum oscillations in BiTeCl with a single frequency and nontrivial Berry phase, and attributed this oscillation to a topological surface state in an earlier report[34]. In this paper, by combining Hall measurements, the Fermi surface tuning, and more

extensive angle-dependent SdH measurements, we find that the oscillations originate from the bulk and that the previous observation of the SdH oscillation with a single frequency corresponds to the extremal case when the Fermi level is located in the vicinity of the Dirac point, where the oscillation from the IFS disappears. When preparing this manuscript, we became aware of another work which also concludes that SdH oscillations in BiTeCl originate from the bulk[35]. Our work is different from Ref. 35, however, as we observe oscillations from both the IFS and the OFS in lower magnetic field, the topological transition of the FS by tuning the carrier density with self-doping effect, the nontrivial $\pi$ Berry phase of both the IFS and OFS. Furthermore we use angle-dependent measurements to resolve the 3D Fermi surface. (2) A topological transition of the FS was also observed in BiTeI, but the relative position of the Fermi level and the band crossing point were tuned by pressure, which modifies the band structure[36].

**Acknowledgement**

This work is supported by the Australian Research Council (ARC) through an ARC Discovery Project (DP130102956. X.L.W) and an ARC Professorial Future Fellowship project (FT130100778, X.L.W).

**Author contributions**

X.L.W. and F.X.X. conceived and designed the experiments. F.X.X. carried out the experiments. F.X.X and X.L.W. analysed the data and wrote the paper. M.V. S.X.D. and M.S.F. contribute to the data analysis and discussions.

**Competing financial interests statement**

The authors claim no competing finical interests.

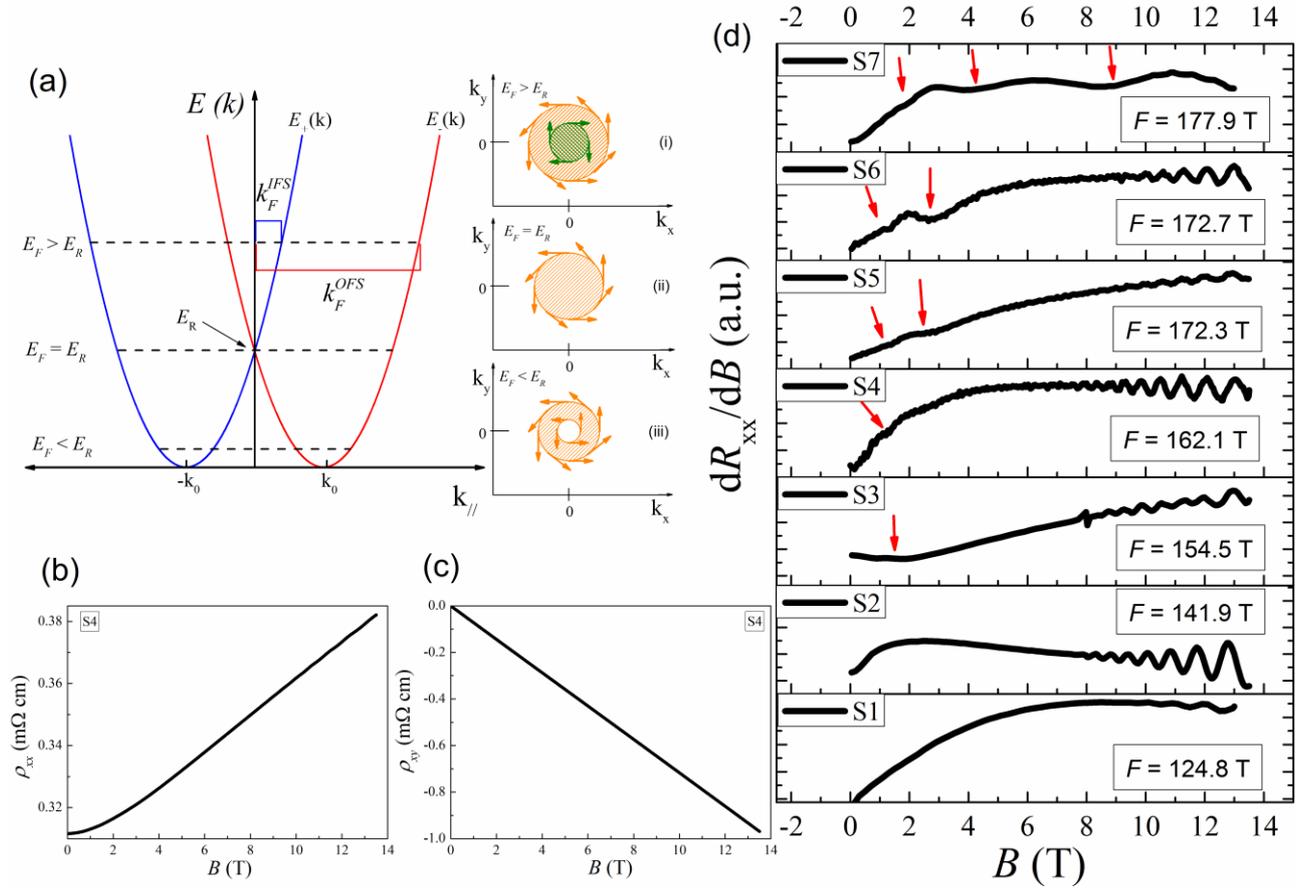

**Figure 1 | Evolution of topological transition of FS.** (**a**) Energy–momentum dispersion in Rashba spin split conduction band. The dashed lines represent different Fermi levels $E_F$. The insets shows constant energy contours for $E_F > E_R$, $E_F = E_R$ and $E_F < E_R$, respectively. In the main Fig. 1a, the blue and red colors indicate two opposite spin direction while in the inset, the blue and red colors indicate different spin helicity and the arrows show the spin direction. (**b**, **c**) Longitudinal resistivty $\rho_{xx}$ and Hall resistivity $\rho_{xy}$ for S4. (**d**) Evolution of the topological transition as the Fermi level shifts from well above the Dirac point down to the vicinity of the Dirac point. The arrows indicate the oscillation periods from IFS that can be observed.

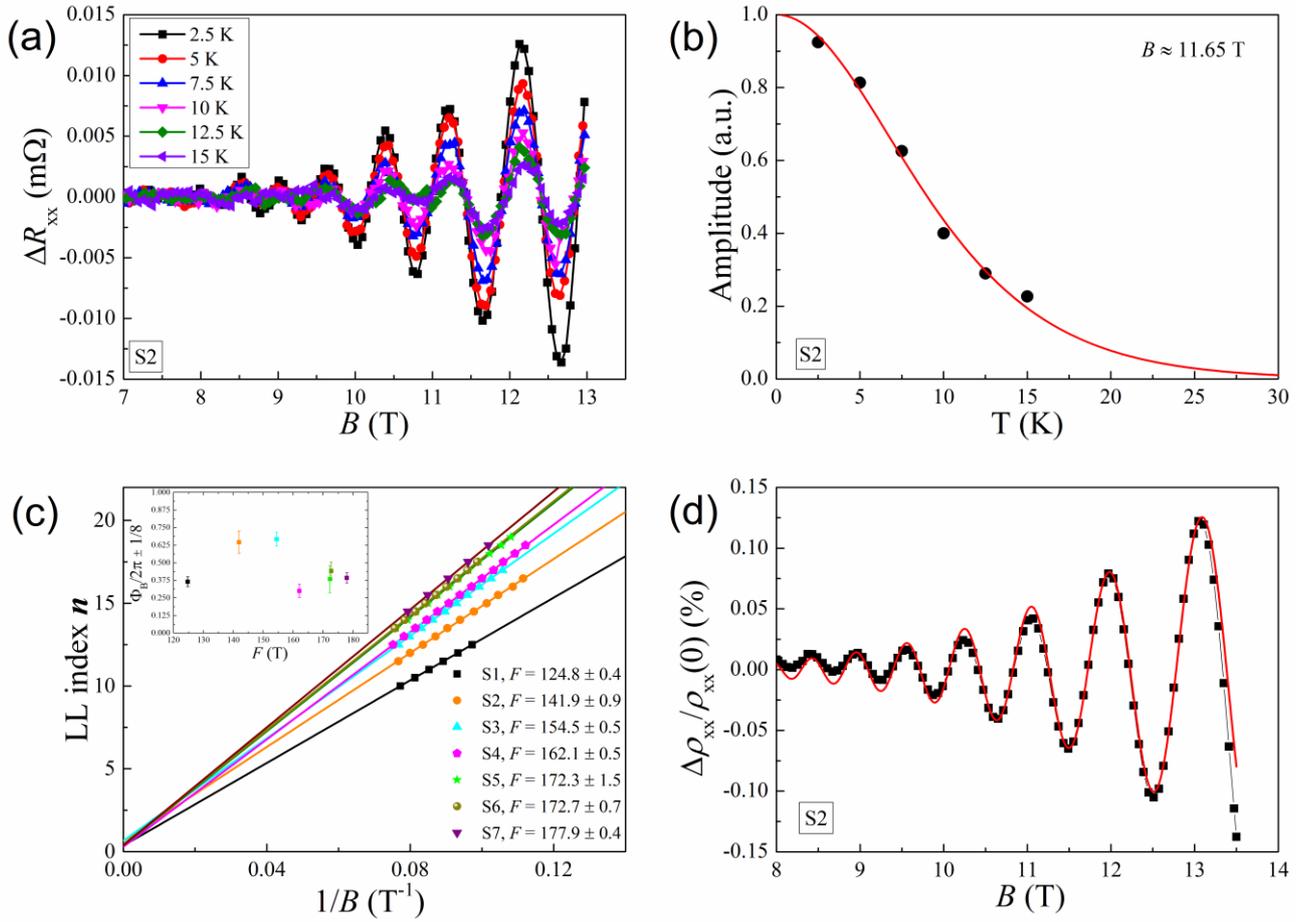

**Figure 2 | Standard SdH oscillation analysis of OFS.** (**a**) OFS SdH oscillations at various temperatures of S2 after subtracting the background. (**b**) Temperature dependence of the OFS oscillation amplitude of S2. Fitting with the thermal damping factor yields the effective mass. (**c**) Landau level fan diagram used to obtain the oscillation frequencies and phase factors of the outer Fermi surfaces for Samples S1-S7. The solid symbols denote LL indices for the minima and maxima of the SdH oscillations. The solid lines are the linear fits to the experimental data, the slopes of which are the frequencies, *F*, of the oscillations given in the legend. The values of the intercepts of the fitting lines with the LL index axis are shown in the inset, and, as shown, most of the intercepts are located in the range of 0.375~0.625, which corresponds to the non-trivial Berry's phase, as discussed in the main text. Each colour corresponds to one sample. The error bars in the inset indicate the standard deviation of the fitting errors. (**d**) OFS SdH oscillations of S2 fitted by the LK formula, which yields the Dingle temperature.

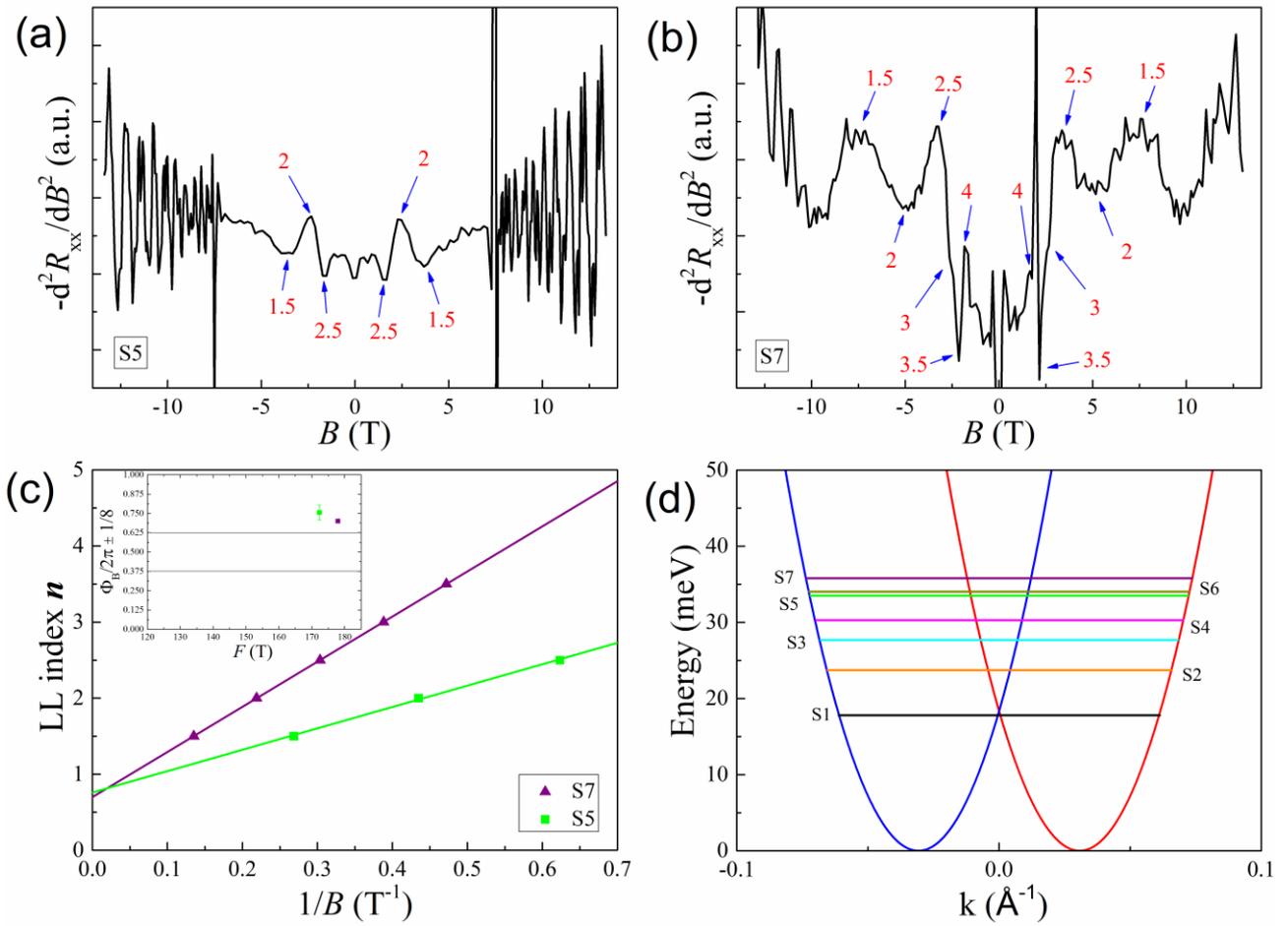

**Figure 3 | SdH oscillation, Berry phase from IFS of S5 and S7, and the band dispersion with the Fermi levels of S1-S7.** $-d^2R_{xx}/dB^2$ as a function of magnetic field for (**a**) S5 and (**b**) S7. The blue arrows indicate the positions of integer and half integer Landau levels. (**c**) Landau level fan diagram for IFS of S3 and S7. The inset shows the values of the intercepts of the fitting lines with the LL index axis which are very close to the region from 0.375 to 0.625 and corresponds to the non-trivial π Berry's phase. The error bars in the inset indicate the standard deviation of the fitting errors. (**d**) The energy-momentum dispersion determined by the Rashba parameters from the experimental results. The blue and red curves represent two spin split bands with opposite spin direction. The solid lines are the Fermi levels for S1-S7 calculated by the Rashba model.

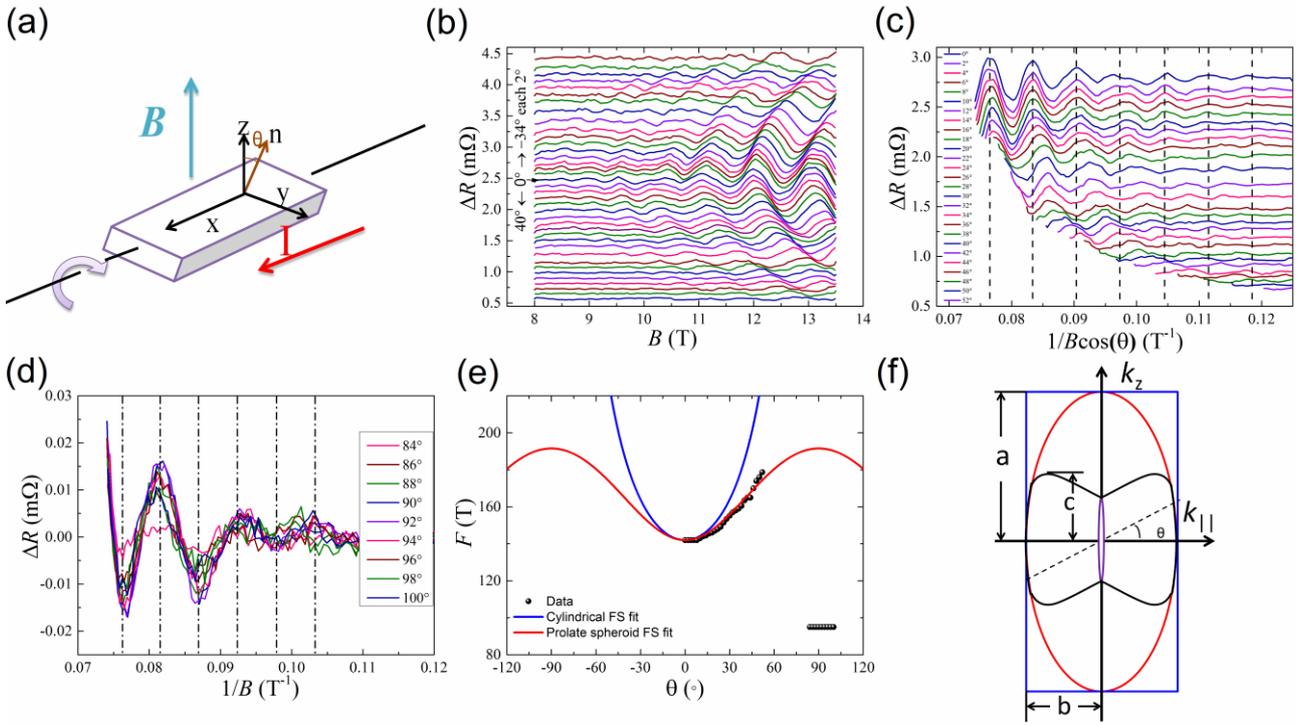

**Figure 4| Mapping the 3D Fermi surface.** (**a**) Measurement configuration for angle-dependent SdH oscillation. The electric current and rotation axis are parallel to the *x* direction, and the magnetic field is along the *z* direction. The tilt angle $\theta$ is defined as the angle between the normal vector of the surface and the *z* axis. (**b**) OFS SdH oscillations of S2 measured from -34° to 40° at 2.5 K. The oscillations are obtained after removing the polynomial background and have been vertically offset for clarity. The observed oscillation periods decrease with increasing tilt angle. (**c**) OFS SdH oscillations of S2 from 0° to 52° in Fig. 4(b) are plotted as functions of $1/B\cos\theta$. The oscillations are obtained after removing the polynomial background and have been vertically offset for clarity. The dash lines indicate the oscillation period at 0° scales with $1/B$, which is consistent with the Landau quantization. (**d**) The OFS SdH oscillations of S2 around 90° reveals its spherical nature. (**e**) Tilt angle dependence of the oscillation frequency. The solid black circle is the experimental data for S2. The blue and red solid lines are the fits of angle dependence of the oscillation frequency from cylindrical ($F_{(\theta)} \propto \frac{1}{\cos\theta}$) and prolate spheroid ($F_{(\theta)} \propto \sqrt{(\frac{a}{b})^2 \cos^2\theta + \sin^2\theta}$) FS respectively. (**f**) Vertical cross-sectional view of the 3D FS cut

along the $k_z$ direction for S2. The blue rectangle denotes the cylindrical FS of 2D electronic system and red ellipse denotes a prolate spheroid FS with major axis a = 119.6 × $10^{-3}$ Å$^{-1}$ and minor axis b = 65.69 × $10^{-3}$ Å$^{-1}$, respectively. The black and purple solids represent the OFS and IFS, which is proposed to explain the observations in Fig. 4(b-e). The dashed line indicates the extremal cross-section of the FS perpendicular to the magnetic field.

Table

| Table 1 \| Parameters determined from SdH oscillations and Rahsba model. $m^{*OFS}$ is the effective mass of the OFS. $F^{OFS}$ and $F^{IFS}$ are oscillation frequencies from the OFS and IFS which yield the Fermi wave vectors, $k_F^{OFS}$ and $k_F^{IFS}$, via the Onsager relation. $T_D^{OFS}$ is the Dingle temperature of the OFS, $E_F$ is the Fermi energy calculated by the Rashba model. $n$ is the carrier density for the 3D model, $(1/3\pi^2)(2eF/\hbar)^{3/2}$. | | | | | | | | |
|---|---|---|---|---|---|---|---|---|
| Sample | $m^{*OFS}$ | $F^{OFS}$ (T) | $T_D^{OFS}$ (K) | $k_F^{OFS}$ ($10^{-3}$ Å) | $\mu^{OFS}$ (cm$^2$V$^{-1}$s$^{-1}$) | $F^{IFS}$ (T) | $k_F^{IFS}$ ($10^3$ Å) | $E_F$ (meV) | $n$ ($10^{18}$ cm$^{-3}$) |
| S1 | 0.203 ± 0.002 | 124.8 ± 0.4 | 22.98 ± 0.19 | 61.59 | 458.5 | | | 18.38 | 7.9 |
| S2 | 0.191 ± 0.005 | 141.9 ± 0.9 | 17.5 ± 0.12 | 65.69 | 640.2 | | | 23.61 | 9.58 |
| S3 | 0.199 ± 0.003 | 154.5 ± 0.5 | 20 | 68.53 | 537.4 | | | 27.61 | 10.88 |
| S4 | 0.194 ± 0.003 | 162.1 ± 0.5 | 20.5 | 70.20 | 537.8 | | | 30.11 | 11.69 |
| S5 | 0.194 ± 0.006 | 172.3 ± 1.5 | 21.05 ± 0.10 | 72.37 | 522.9 | 2.81 ± 0.10 | 9.24 | 33.52 | 12.81 |
| S6 | 0.197 ± 0.002 | 172.7 ± 0.7 | 15.53 ± 0.20 | 72.45 | 699.1 | | | 33.65 | 12.86 |
| S7 | | 177.9 ± 0.4 | | 73.54 | | 5.93 ± 0.01 | 13.43 | 35.44 | 13.45 |

# Supplementary Information for

# Observation of topological transition of Fermi surface from a spindle-torus to a torus in large bulk Rashba spin-split BiTeCl

Fei-Xiang Xiang[1], Xiao-Lin Wang[1]*, Menno Veldhorst[2], Shi-Xue Dou[1], Michael S. Fuhrer[3]

[1]*Institute for Superconducting and Electronic Materials, Australian Institute for Innovative Materials, University of Wollongong, Innovation Campus, North Wollongong, New South Wales 2500, Australia,*

[2]*Centre for Quantum Computation and Communication Technology, School of Electrical Engineering and Telecommunications, The University of New South Wales, Sydney, New South Wales 2052, Australia,*

[3]*School of Physics, Monash University, Clayton, Victoria 3800, Australia*

*Corresponding author: xiaolin@uow.edu.au*

**This file includes:**

**Part *A*: Typical SdH oscillations without background subtraction.**

**Part *B*: Standard SdH oscillation analysis with LK formula on more samples.**

**Part *C*: Angle-dependent SdH oscillation frequency of a prolate spheroid FS.**

**Part *D*: Angle dependence of SdH oscillations for S4.**

**Part *E*: Assignment of Landau level index and determination of Berry phase.**

**Part *F*: Extraction of SdH oscillation frequency with fast Fourier transform.**

**Part *A*: Typical SdH oscillations without background subtraction.**

Figure 1d,e shows two typical Shubnikov-de Haas oscillations without background subtraction observed in measurements. In Fig. 1d clear SdH oscillations can only be observed in high magnetic field as indicated by the blue arrow, while in Fig. 1e weak SdH oscillations can be observed in both low and high magnetic field as indicated by the blue and red arrows, respectively.

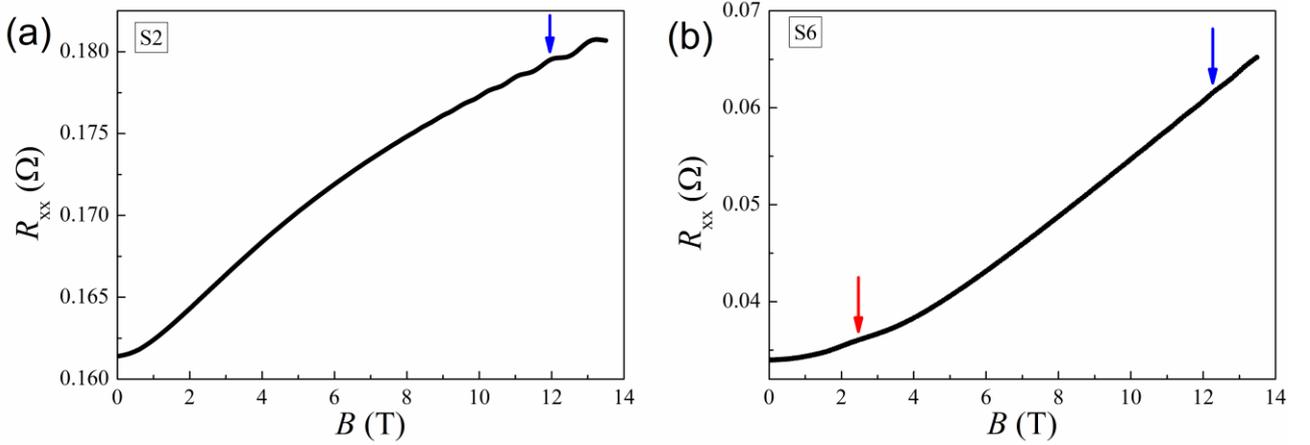

**Figure S1**: (**a**, **b**) Typical SdH oscillations without background subtraction.

**Part *B*: Standard SdH oscillation analysis with LK formula on quantum oscillations of more samples.**

In addition to the standard SdH oscillation analysis with LK formula for quantum oscillations from OFS of S2 in Fig. 1(a)-(c), the same analysis method used in the main text was used for other samples except for S7 as shown in Fig. S1-3. Combing with the LL fan diagram, effective mass $m_e$, oscillation frequency, $F$, and dingle temperature, $T_D$, can be extracted as shown in the table 1 in main text.

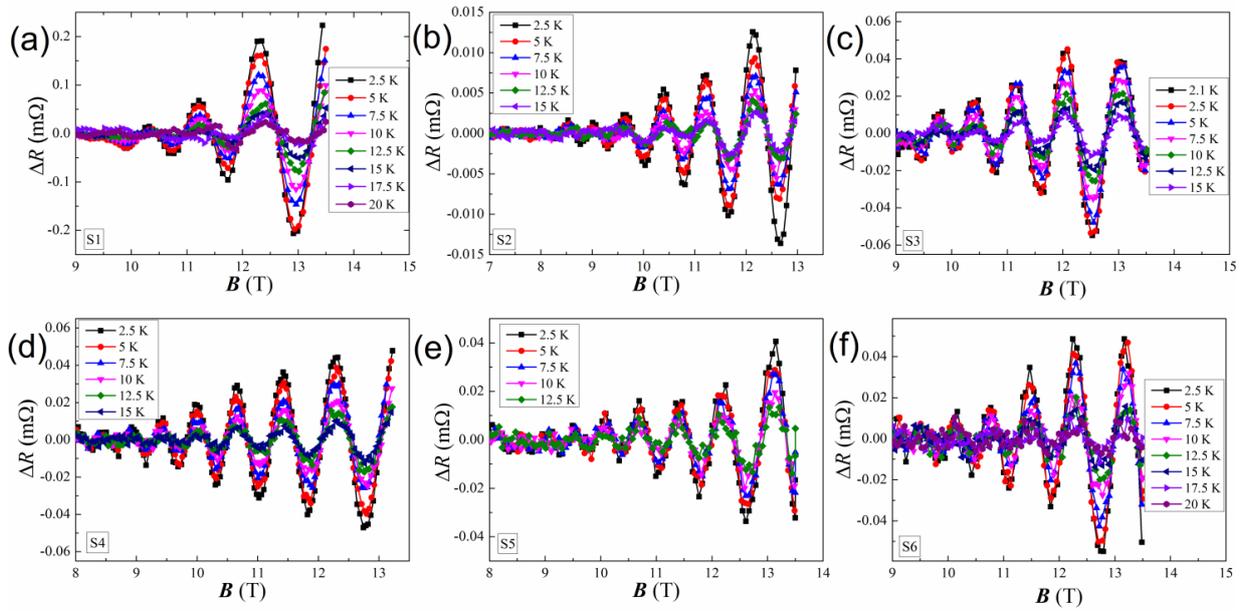

**Figure S2**: (**a**-**f**) SdH oscillations measured at various temperatures for S1-S6. The oscillations are obtained after removing the background.

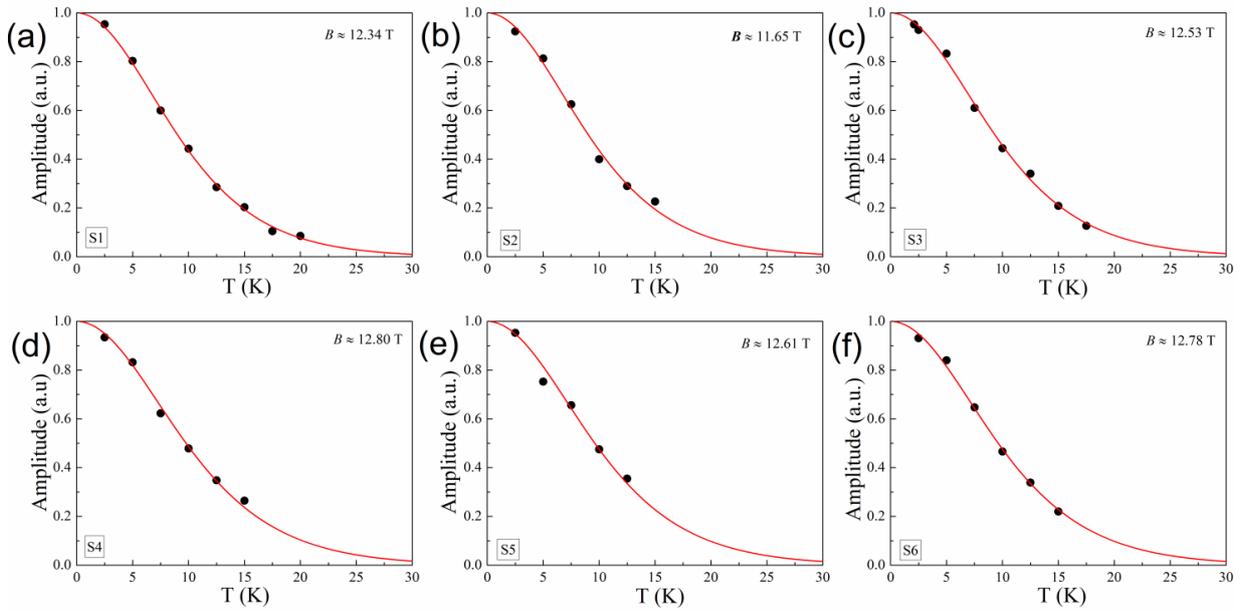

**Figure S3**: (**a**-**f**) Temperature dependence of the oscillation amplitude for S1-S6. The red solid lines are the fitting with thermal damping factor, $\frac{2\pi^2 k_B T m^*/\hbar eB}{\sinh(2\pi^2 k_B T m^*/\hbar eB)}$, which yields the effective mass.

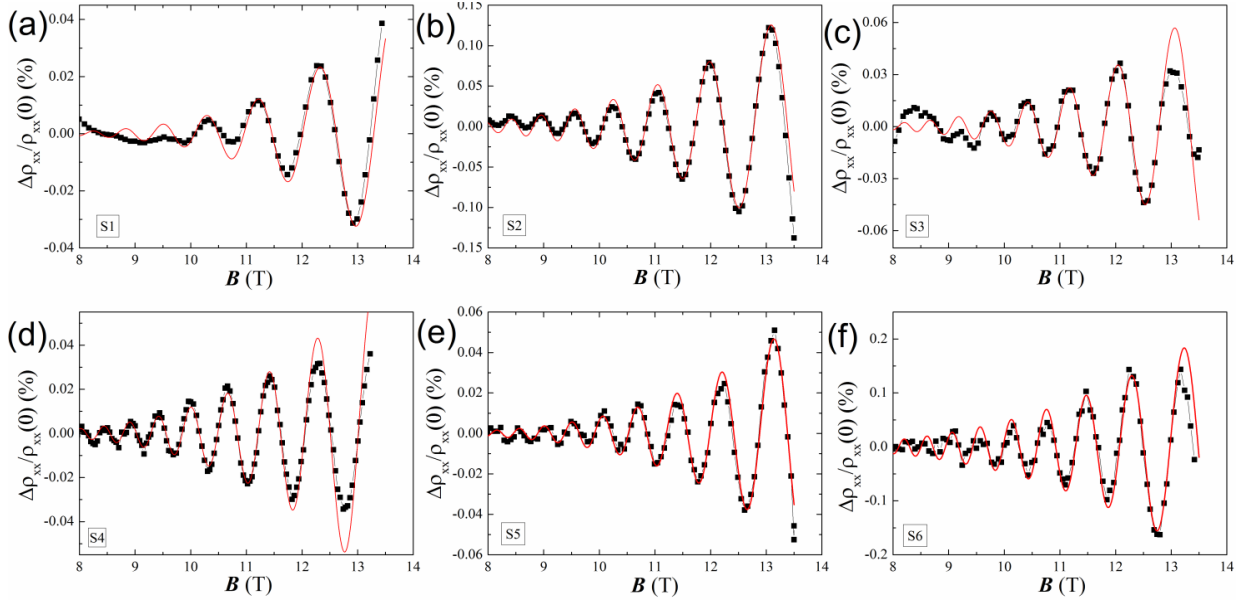

**Figure S4**: (**a-f**) LK fitting of SdH oscillations at 2.5 K for S1-S6, which yields the dingle temperatures.

**Part *C*: Angle-dependent SdH oscillation frequency of a prolate spheroid FS.**

According to the angle-dependent SdH oscillation in Fig. 4c, the FS deviates from cylindrical shape for two-dimensional electronic system. To describe the 3D FS in BiTeCl, several FS geometries were tried and it was found that a prolate spheroid based FS can fit reasonable well the angle-dependent oscillation frequency when $\theta \leqslant 52°$. The equation of the prolate spheroid, in which two semi-axes with length, b, in the $k_{//}$ plane and one semi-axis, a, in the $k_z$ direction can be expressed as:

$$\frac{x^2+y^2}{b^2} + \frac{z^2}{a^2} = 1, (a > b)$$

The extremal cross section of the FS perpendicular to the magnetic field is an ellipse ($\theta > 0$) and a circle ($\theta = 0$), its equation can be expressed as:

$$\frac{x^2}{b^2} + \frac{y^2}{a^2 \sin^2\theta + b^2 \cos^2\theta} = 1, (\theta \text{ is the tilt angle})$$

Therefore, the extremal cross section area of the FS perpendicular the magnetic field is

$$A_F = \pi b^2 \sqrt{\left(\frac{a}{b}\right)^2 \sin^2\theta + \cos^2\theta}$$

Because $F = (\hbar/2\pi e)A_F$, the $F$ as a function of $\theta$ can be expressed as

$$F = \frac{\hbar b^2}{2e} \sqrt{\left(\frac{a}{b}\right)^2 \sin^2\theta + \cos^2\theta}$$

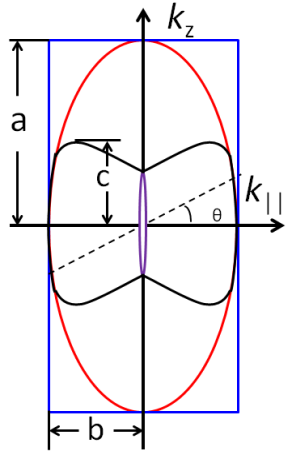

**Figure S5**: Vertical cross-sectional view of the 3D FS cut along the $k_z$ direction for S2. The blue rectangle denotes the cylindrical FS of 2D electronic system and red ellipse denotes a prolate spheroid FS with major axis a = 119.6 × 10$^{-3}$ Å$^{-1}$ and minor axis b = 65.69 × 10$^{-3}$ Å$^{-1}$, respectively. The black and purple solids represent the OFS and IFS, which is proposed to explain the observations in Fig. 4(b-e). The dashed line indicates the extremal cross-section of the FS perpendicular to the magnetic field.

**Part *D*: Angle dependence of SdH oscillations from OFS of S4.**

Besides the angle dependence of SdH oscillation analysis in sample S2, similar measurements were also carried out in sample S4. Fig. S4(a) shows the quantum oscillations from -6° to 46°, in which as the increase of tilt angle the number of the observed oscillation period become less and the oscillation amplitude diminishes, similar to S2. Fig. S4(b) shows the same quantum oscillations from Fig. S4(a) in 1/*B* scale. The dash lines in Fig. 4(b) indicate the period of oscillations at 0°. It can be seen clearer that the oscillation from OFS of S4 exhibits stronger two-dimensional feature compared with those of S2, which agree with the topological transition of FS. Because the Fermi level of S4 is higher than that of S2, the Fermi surface are more like spindle-torus shape and OFS are more close to cylinder shape.

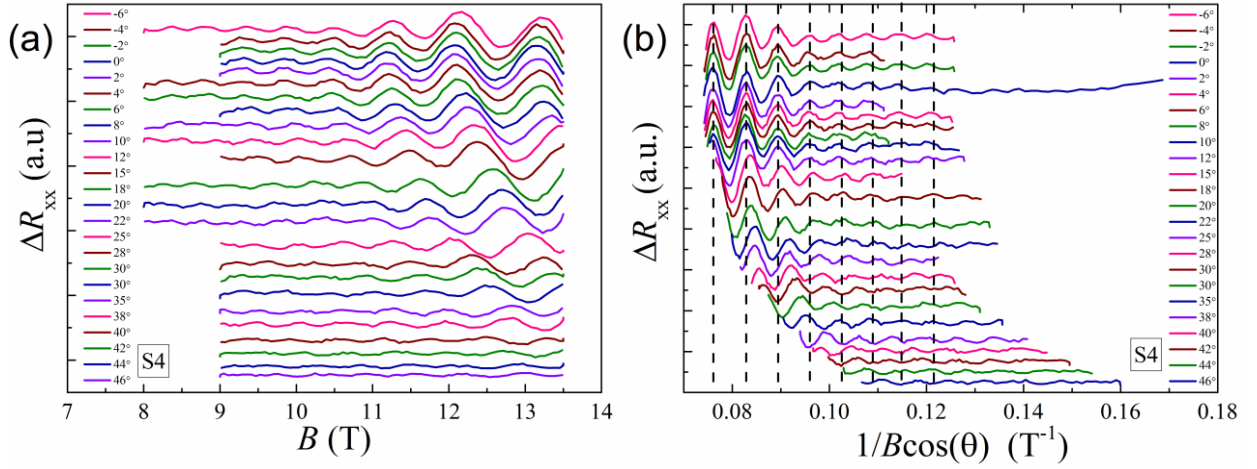

**Figure S6**: (**a**) SdH oscillations from OFS of S4 at various angles after removing the background. The curves are vertically offset for clarity. (**b**) SdH oscillations in Fig. S4(**a**) are plotted in $1/B$ scale. The dash lines indicate the oscillation period.

**Part *E*: Assignment of Landau level index and determination of Berry phase.**

The measurement of both longitudinal resistivity and Hall resistivity is important not only for calculating the carrier density and Hall mobility, but also for assigning of the Landau level (LL) index and determining the Berry phase. Because of $\sigma_{xx} = \frac{\rho_{xx}}{\rho_{xx}^2 + \rho_{xy}^2}$, while for $\rho_{xx} \gg \rho_{xy}$, the minima and maxima of the oscillations in longitudinal conductivity $\sigma_{xx}$ are out of phase with those in longitudinal $\rho_{xx}$, integer LL ***n*** is assigned to oscillatory maxima of $\rho_{xx}$; for the case of $\rho_{xx} \ll \rho_{xy}$, the minima and maxima of the oscillations in $\sigma_{xx}$ are in phase with those in $\rho_{xx}$, integer LL ***n*** is assigned to oscillatory minima of $\rho_{xx}$. However, how the phase factor shifts when $\rho_{xx} \sim \rho_{xy}$ is still unclear. To answer this question, we establish a simple model to estimate the phase shift window. Assumption 1: $\rho_{xx} = \Delta \rho_{xx} + \rho_{xx}^{const}$, $\Delta \rho_{xx} \ll \rho_{xx}^{const}$, where the $\Delta \rho_{xx}$ is the oscillatory part of $\rho_{xx}$ and $\rho_{xx}^{const}$ is the background and assumed to be a constant for simplicity. (In the our case as shown in Fig. S3, $\frac{\Delta\rho_{xx}}{\rho_{xx}^{const}} \sim \frac{\Delta\rho_{xx}}{\rho_{xx}} < \frac{1}{1000}$)

Assumption 2: The oscillations in the experiment come from $\Delta\rho_{xx}$.

Now when $\rho_{xx} \ll \rho_{xy}$, which is always the case in a one-band model when SdH oscillation is observed ($\rho_{xy} = \rho_{xx} * \mu B$, with $\mu B \gg 1$ in order to have SdH oscillations). Then we can write:

$$\sigma_{xx} = \frac{\rho_{xx}}{\rho_{xx}^2 + \rho_{xy}^2} \sim \frac{\rho_{xx}^{cont}}{\rho_{xy}^2} + \frac{\Delta\rho_{xx}}{\rho_{xy}^2}$$

So $\sigma_{xx}$ is in phase with $\rho_{xx}$.

When $\rho_{xx} \gg \rho_{xy}$ (which is possible in a multiband case where there is a dominant low mobility band present), then we can write:

$$\sigma_{xx} = \frac{\rho_{xx}}{\rho_{xx}^2 + \rho_{xy}^2} \sim \frac{1}{\rho_{xx}}$$

So $\sigma_{xx}$ is out of phase with $\rho_{xx}$.

In the case of $\rho_{xx} \sim \rho_{xy}$, because of $\frac{\Delta\rho_{xx}}{\rho_{xx}^{const}} < \frac{1}{1000}$ as shown in Fig. S3, $\rho_{xx}$ can be written approximately as $\rho_{xx} = 1000 + \cos(50/B)$, where the oscillation amplitude is assume to be 1 for simplicity and the constant part is set to 1000. As $\rho_{xy}$ is varied from 2 to 50,000, $\sigma_{xx}$ is plotted as a function of magnetic field in reciprocal scale as shown in the Fig. S5. For $\rho_{xx} \ll \rho_{xy}$, ie $\rho_{xy} = 5000$, and $\rho_{xx} \gg \rho_{xy}$, ie $\rho_{xy} = 2$, the phase factors in $\sigma_{xx}$ differ 180°. When $\rho_{xx} \sim \rho_{xy}$, it was found the phase changes just in a very narrow window. The change in $\rho_{xy}$ on the order of $2 * \Delta\rho_{xx}$, which equals to $2\cos(50/B) \sim 2$ in this case, will shift the phase by 180°. When $999 \leq \rho_{xy} \leq 1001$, the period of oscillation become half.

As experimental data shown in Fig. 1b,c the regime of $\rho_{xx} \sim \rho_{xy}$ is around 3~5 T. For the SdH oscillations from OFS of S1 to S7, $\rho_{xx} < \rho_{xy}$, so the integer LL *n* is assigned to the minima of SdH oscillations. For the SdH oscillations from IFS of S5 they can be categorized into the regime of $\rho_{xx} > \rho_{xy}$, so the integer LL $n$ is assigned to the maxima of SdH oscillations as shown in Fig. S6(a). For the SdH oscillations from IFS of S7, because the oscillations extend from the low field, around 1.5 T, to high field, around 8 T, so in the low field part the integer LL $n$ is assigned to the minima of SdH oscillations and in the high field part the integer LL $n$ is assigned to the maxima of SdH oscillations as shown in Fig. S6(b). The dash dot lines indicate the minima and maxima of oscillations and also indicate that the oscillatory periods are in $1/B$ scale which agrees with the Landau quantization.

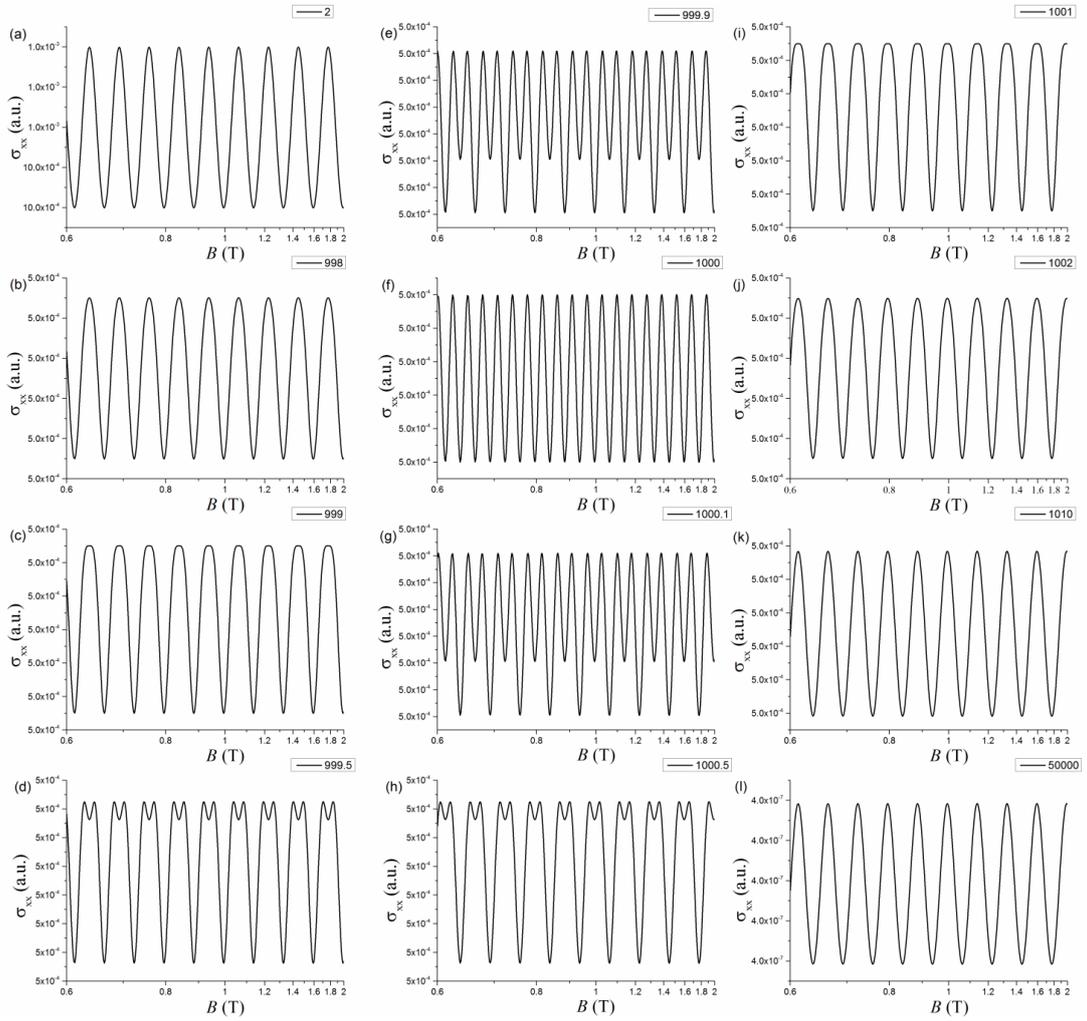

**Figure S7**. (**a-i**) Longitudinal conductivity, $\sigma_{xx}$, as a function of magnetic field in reciprocal scale is plotted for various $\rho_{xy}$. Among them $\rho_{xy}$ = 5000 and 2 correspond to $\rho_{xx} \ll \rho_{xy}$ and $\rho_{xx} \gg \rho_{xy}$, respectively.

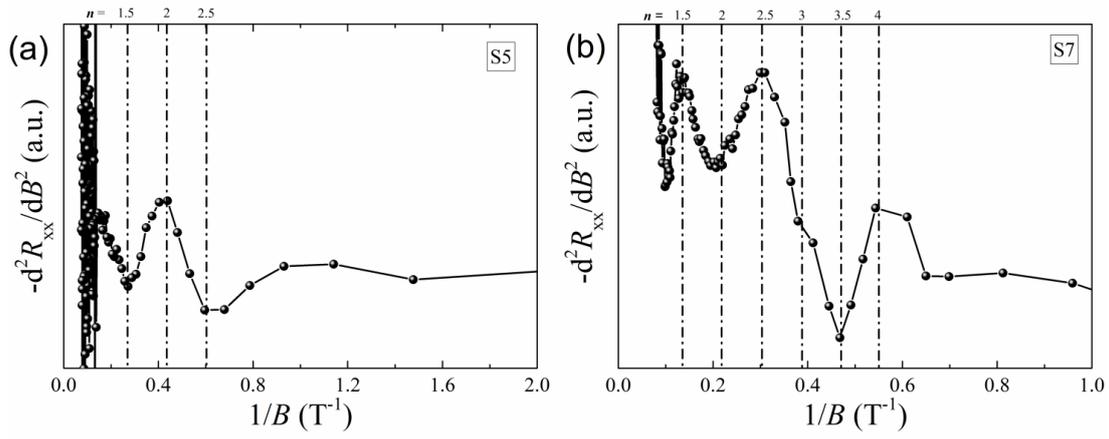

**Figure S8**. – d²$R_{xx}$/d$B^2$ as a function of 1/$B$ for sample S5 and S7. The dash dot lines mark the minima and maxima of SdH oscillations indicating the oscillatory periods are in 1/$B$ scale.

**Part *F*: Extraction of SdH oscillation frequency with fast Fourier transform.**

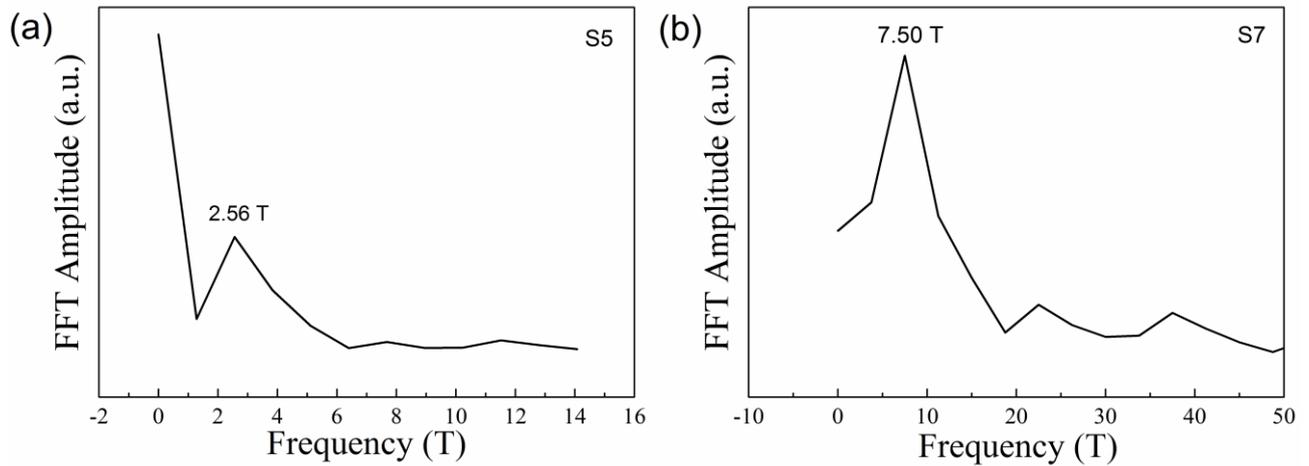

Figure S9, Fast Fourier transform spectrum of SdH oscillation frequency.

Figure S9 shows the SdH oscillation frequencies for S5 and S7 yielded from fast Fourier transform. They are quite close to the oscillation extracted from Landau level fan diagram analysis, 2.81 T and 5.93 T for S5 and S7, respectively.